\documentclass[preprint,authoryear,12pt]{article}

\usepackage{geometry,float}
\usepackage{amssymb}
\usepackage{amstext}
\usepackage{bm}
\usepackage{graphics}
\usepackage{graphicx}
\usepackage{amsthm,amsfonts,epic,latexsym}
\usepackage{amsmath}
\usepackage{latexsym}
\usepackage{rotate}
\usepackage{lscape}
\usepackage{color}
\usepackage{epsfig,epsf,psfrag}
\usepackage{sectsty}
\usepackage{graphics}
\usepackage{epsfig}
\usepackage{multirow}
\usepackage{graphicx}
\usepackage{sectsty}
\usepackage{epsfig,natbib}
\usepackage{rotate}
\usepackage{lscape}
\usepackage{graphicx}
\usepackage{enumitem}
\usepackage{setspace}
\def\bco{\iffalse}
\definecolor{DarkRed}{rgb}{.7,0,.4}
\def\red{\textcolor{DarkRed}}

\include{mathdefn2}

\newcommand{\bea}{\begin{eqnarray*}}
\newcommand{\eea}{\end{eqnarray*}}
\newcommand{\be}{\begin{eqnarray}}
\newcommand{\ee}{\end{eqnarray}}
\newcommand{\ed}{\end{document}}

\newcommand{\btab}{\begin{tabular}}
\newcommand{\etab}{\end{tabular}}

\newcommand{\bi}{\begin{itemize}}
\newcommand{\ei}{\end{itemize}}
\newcommand{\bfi}{\begin{figure}}
\newcommand{\efi}{\end{figure}}
\newcommand{\ben}{\begin{enumerate}}
\newcommand{\een}{\end{enumerate}}
\newcommand{\bay}{\begin{array}}
\newcommand{\eay}{\end{array}}

\def\bco{\iffalse}

\def\ci{\cite}
\def\cp{\citep}

\newcommand{\bc}{\begin{center}}
\newcommand{\ec}{\end{center}}

\definecolor{DarkRed}{rgb}{.7,0,.4}
\def\red{\textcolor{DarkRed}}

\DeclareMathOperator*{\argmax}{argmax}
\DeclareMathOperator*{\argmin}{argmin}

\bibpunct{(}{)}{;}{a}{}{,}

\newcommand{\e}{\varepsilon}

\newcommand{\R}{\mathbb{R}}
\newcommand{\1}[1]{1\!\mathrm{l}\{ #1\}}         
\newcommand{\E}{\mathbb{E}}                       
\newcommand{\EE}{\mathcal{E}}                       
\newcommand{\T}{\mathcal{T}}                      
\newcommand{\C}{\mathcal{C}}                      
\newcommand{\W}{\mathcal{W}}                      
\def\argmin{\mathop{\rm arg\,min}}

\graphicspath{{./Fig/}}

\begin{document}

\title{Pairwise Dynamic Time Warping for Event Data}
\author{Ana Arribas-Gil\\Departamento de Estad\'istica\\
        Universidad Carlos III de Madrid, Getafe, Spain.\\ E-mail: aarribas@est-econ.uc3m.es \and Hans-Georg M\"uller\\Department of Statistics, University of California, Davis\\
Davis, CA 95616 USA}

\date{}
\maketitle

\thispagestyle{empty}

\begin{abstract}
We introduce a new version of dynamic time warping for samples of observed
event times that are modeled as time-warped intensity processes.
Our approach is developed within a framework where
for each experimental unit or subject in a sample, one observes a random number of event times
or random locations. As in our setting the number of observed events differs from
subject to subject, usual landmark alignment methods that require the number
of events to be the same across subjects are not feasible. We address this
challenge by applying dynamic time warping,
initially by aligning the event times for pairs of subjects, regardless of
whether the numbers of observed events within the considered pair of subjects
match. The information about pairwise alignments is then combined to extract
an overall alignment of the events for each subject across the entire sample. This overall alignment provides a useful description of
event data and can be used as a pre-processing step for
subsequent analysis. The method is illustrated with a historical fertility study and with
on-line auction data.
\end{abstract}


\section{Introduction}
Time warping is commonly used in functional data analysis to address the
presence of random variation in time in addition to amplitude variation. A
major difficulty is the non-identifiability of these two components
\cp{mull:04:4}. This non-identifiability requires that one makes strong
assumptions on the nature of the warping mechanism in order to be able to distinguish it from
the variation in the random amplitudes. While various solutions have emerged over
the years, a gold standard for functional time warping is the landmark
method \cp{knei:92, gass:95}. In this method, one typically extracts for each
random curve $X_i$ in a sample of i.i.d. trajectories $X_1, \ldots, X_n$ a
series of landmarks, which correspond to the times or locations where certain
features occur, such as the location of a zero or the location of a peak.

One then uses these locations to represent the sample or for subsequent
continuous time warping of the random functions,
by moving the landmark
locations to their cross-sample average locations. Then one can smoothly map the
time intervals in between these locations, for example by applying  monotone
spline transformations.  Usually, the landmarks are
obtained by presmoothing the often noisily observed functions $X_i$,
obtaining estimates $\hat{X}_i$ and then extracting features such as peak
locations by substituting the unknown peak location $\theta_{i}=\argmax_x
X_i(x)$ by the estimated peak location $\hat{\theta}_{i}=\argmax_x
\hat{X}_i(x),$ where a peak is just one of many
possible features defining a landmark \cp{mull:84:2}.

A persistent difficulty with the landmark method has been that in real data
applications, landmarks such as peaks can be hard to identify for specific
sample curves, due to either noise in the measurements or to the presence
of non-standard trajectories, which do not exhibit some of the landmarks or
have more landmarks. In growth curves, for example,  one might encounter more
than one mid-growth spurt and it is then hard to decide how to position extra
peaks within the sequence of landmarks. The landmark method however requires
that the number of identified landmarks matches across all curves and also
that the landmarks are in a sequence that is equally interpretable across all subjects
in terms of
features (location of peaks and of zeros and size of peaks for functions and
derivatives). In addition, although seemingly straightforward, the
computational task of landmark extraction is actually quite burdensome and
usually cannot be fully automated, so that identifying and verifying a large
number of landmarks even for modest samples of curves can be a major effort
that includes some subjective elements \cp{mull:84:5}.

For these reasons, non-landmark based methods for function warping have met
with increasing interest in recent years in the functional data analysis
literature \cp{silv:95, wang:97,Aach,ronn:01,gerv:04,knei:08, tele:08}. As already
mentioned, such methods depend explicitly or implicitly on strong assumptions
to circumvent the non-identifiability problem.  In this paper, we take a
different approach. We assume from the beginning that the number of landmarks
that can be identified in sample trajectories is random, but that
nevertheless their order matters, so that the lower ranked landmarks should
be placed closer together in time than the higher ranked landmarks, irrespective of the number
of landmarks that are available for each subject.

More generally, the times at which landmarks occur are viewed as random event
times that are observed for each subject. These event times carry information about 
their order and relative location to each other, but otherwise are not distinguishable. 
Typical examples for such event
data that we use to illustrate our methods are age at child birth per woman in a large
French-Canadian historical cohort, where women typically have a large number
of offspring, and the timings of bids submitted at online e-bay
auctions, recorded for each auction in a sample of auctions. Such data can be viewed as realizations of
random intensities of underlying point processes. Related methods for time
synchronization for densities have been studied in \ci{mull:11:4}; for 
general functional methodology for intensities we refer to 
\ci{bouz:06:2, mull:12:2}.

To model such data, denote the $n_i$ observed event times for the $i$-th subject by $t_{i1}, \ldots, t_{in_i}$, and assume that 
all event times are observed within the interval $\T=[0,T]$. We consider standardized cumulative incidence functions $$Y_i(t) = \frac{1}{n_i}\sum_{j=1}^{n_i}
1_{\{t_{ij} \le t\}}, \quad t \in \T, \quad i=1,\ldots,n,$$ and assume these are generated 
by an underlying monotone increasing  
fixed function $\mu$ and warping functions $h_i$, which  are
i.i.d. realizations of a random variable $H$ taking values on $\W(\T)=\{f\in
\C(\T) \mid  f\mbox{ strictly increasing},\, f(0)=0, f(T)=T\}$, such that 
\begin{equation}\label{tw}
Y_i(t_{ij})= \mu(h^{-1}_i(t_{ij})), \quad t_{ij}\in {\cal T}, \quad i=1,\ldots,n, \quad j=1, \ldots,n_i. \end{equation}

Here, $\mu(t)$ is a cumulative distribution function and the constraints on $H$ ensure that 
$\mu(h^{-1}_i(t))$ also are (random) cumulative distribution functions. 
In this framework, registration procedures are defined by estimates of the registration 
functions $h_i$. The determination of the time warping functions $h_i$ is of intrinsic interest and 
also allows to determine unwarped versions $X_i(t)=Y_i(h_i(t))$, $t\in \T$ of the $Y_i$. 
For identifiability purposes, it is expedient to assume that $\E[H(t)]=I_{\cal T}(t)$,
where $I_{\cal T}$ is the identity function on the domain ${\cal T}$.

\section{Methodology}

\subsection{Pairwise registration}

For pairwise registration, we adopt  the approach proposed by \cite{TangMuller_Bio}. The idea is to
obtain global alignment for a sample of random functions, from information about
pairwise alignments. Using this device, the task of global  registration is reduced to the often more manageable  task of constructing pairwise or ``relative'' alignments between pairs of random trajectories.

The initial goal is thus to obtain pairwise synchronizations for all or many  
pairs of curves $Y_i$ and $Y_j$, as defined in (\ref{tw}).  For this purpose, 
\cite{TangMuller_Bio} define 
pairwise synchronization functions
\begin{equation}\label{g}
g_{ji}(t)=h_j(h_i^{-1}(t)),
\end{equation}
which provide a transformation of the time scale of the curve $Y_j$ towards that of $Y_i$, and show that $\E[h_j(h_i^{-1}(t))|h_i^{-1}(t)]=h_i^{-1}(t)$. This motivates the estimators 
\begin{equation}
\hat{h}_i^{-1}(t)=\dfrac{1}{n-1} \sum_{j\neq i}  \hat{g}_{ji}(t).
\end{equation}

The alignment problem is then reduced to the problem of defining appropriate estimators $\hat{g}_{ji}(t)$, that
can take into account the particular nature of the data, in our case event data. While 
 \cite{TangMuller_Bio} proposed minimization of a penalized $L^2$-distance to obtain 
appropriate estimates of $g_{ji}(t)$, in the situation we study here, dynamic time warping (DTW) 
provides a promising alternative.

\subsection{Dynamic time warping}

Dynamic time warping (DTW) is a series of alignment algorithms
originally developed in the 1970s in the context of speech
recognition (see \cite{KruskalLib} for an overview).
These are dynamic programming algorithms that, given a
certain metric, consist on aligning two sequences of outcomes by
minimizing the total distance between them, computed
as the sum of distances between each pair of points
along the aligned positions in a suitable metric. Dynamic time warping is
similar to the algorithms used for the alignment of 
biological sequences, such as the Needleman-Wunsch
algorithm \citep{NW} for global alignment.  

\begin{figure}[!h]
\fbox{{\parbox{0.98\linewidth}{
\vspace{0.5cm}
\begin{minipage}{7.25cm}
\begin{center}
\setlength{\unitlength}{0.7mm}
\begin{picture}(45,45)(10,0)
\put(6,6){\line(1,0){56}} \put(6,6){\line(0,1){46}}
\multiput(16,6)(10,0){5}{\dashbox{1}(0,46)}
\multiput(6,16)(0,10){4}{\dashbox{1}(56,0)}
\put(9,-1){\Large $a_1$}\put(19,-1){\Large $a_2$}\put(29,-1){\Large
$a_3$}\put(39,-1){\Large $a_4$} \put(49,-1){\Large $a_5$} \put(-1,9){\Large $b_1$}\put(-1,19){\Large
$b_2$}\put(-1,29){\Large $b_3$}\put(-1,39){\Large $b_4$}
\thicklines
\drawline[5](6,6)(16,16)(26,16)(36,16)(46,26)(56,36)(56,46)
\linethickness{0.35mm}
\put(16,16){\line(1,0){10}}\put(26,16){\line(1,0){10}}\put(56,36){\line(0,1){10}}
\end{picture}\vspace{0.35cm}\\
Alignment:\vspace{-0.1cm}
$$\e=\{ (1,1) , (1,0) , (1,0) ,(1,1), (1,1) ,(0,1)\} $$
\end{center}
\end{minipage}
\begin{minipage}{8.6cm}
\bco Biological sequences alignment interpretation:
$$\begin{array}{ccccccc}
a_1 & a_2 & a_3 & a_4 & a_5 & - &\\
 |  &     &     &  |  &  |  &    &\\
b_1 & -   & -   & b_2 & b_3 & b_4&
\end{array}$$ \fi
Dynamic time-warping interpretation:
$$\begin{array}{cccc}
           a_1  \,  a_2 \, a_3 & a_4  &   a_5 &\\
     \backslash \, | \, /      &  |   &   / \, \backslash      &\\
                b_1            & b_2  &  b_3 \, b_4&
\end{array}$$\vspace{0.25cm}\quad
\end{minipage}
}}}
\caption{Graphical representation of a dynamical time warping alignment between two sequences
  ${\bf a}= (a_1, \ldots, a_{\ell})$ and ${\bf b}= (b_1, \ldots, b_m)$.
   Lines represent correspondence between
   elements of the two sequences.}\label{alig}
\end{figure}
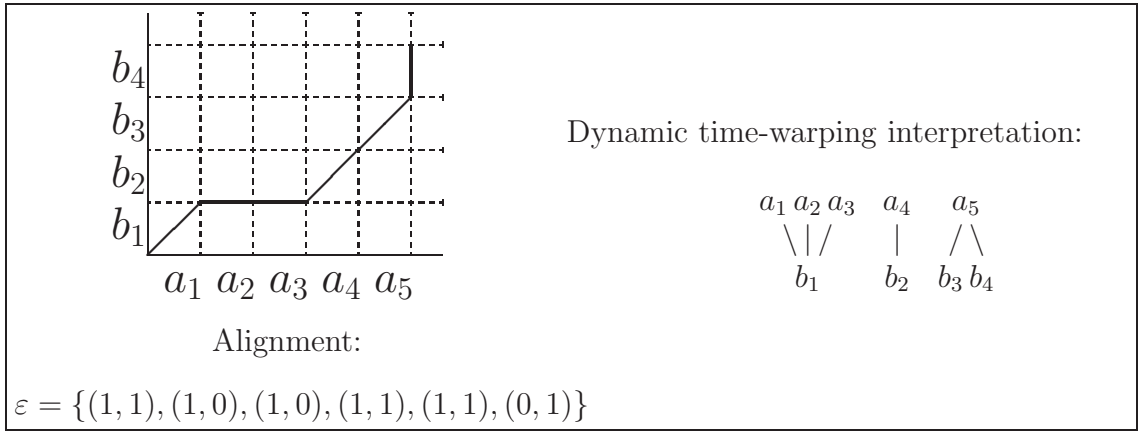

In this context, given two sequences
${\bf a}= (a_1, \ldots, a_{\ell})$ and ${\bf b}=(b_1,\ldots,b_m)$, with values on
a certain feature space ${\cal A}$, an alignment of ${\bf a}$ and ${\bf b}$
is a sequence $\{\e_k\}_{k=1}^{|\e|}$ of variable length $|\e|$, taking
values on $\{(0,1), (1,0), (1,1)\}$ such that $\sum_{k=1}^{|\e|}\e_k=({\ell},m)$.
That is, $(L_r,M_r)=\sum_{k=1}^{r}\e_k$, $r=1,\ldots,|\e|$, defines a path
from $(0,0)$ to $({\ell},m)$ in the two-dimensional integer lattice. We say that
$a_i$ is aligned to $b_j$ if there exists a $1\leq r\leq |\e|$ such that
$\sum_{k=1}^r \e_k=(i,j)$. \bco In fact, this is slightly different to the
biological sequences alignment framework in which we say that two positions
$a_i$ and $b_j$ are aligned to each other if it exists $1\leq r\leq |\e|$
such that $\sum_{k=1}^r \e_k=(i,j)$ and $\e_r=(1,1)$. The difference relies
on the fact that \red{I think the biological sequence discussion is an unnecessary 
distraction}. \fi Thus in dynamic time warping one allows one element in one
sequence to be aligned to more than one element in the other sequence, which makes this method suitable for sequences of event times with unequal number of events in each sequence. The general approach is illustrated in Figure \ref{alig}.
\bco and
all the elements of both sequences must have a correspondent in the other one.
Whereas in the biological sequences alignment framework each element may only
be aligned to one element of the other sequence and some elements may not be
aligned to anything, given place to which is known as insertions and
deletions in the sequence (see \cite{Durbin} for an overview on biological
sequences alignment). This difference is illustrated in Figure \ref{alig}. In
fact, generally speaking, the last approach is the usual one for sequence
comparison when sequences take values on a finite non numerical alphabet, and
dynamic time-warping is the benchmark for numerical continuous feature space synchronization, in which expansions and contractions are the analogous to
insertions and deletions.

Let us point out another important difference: whereas in biological sequence alignment
the time space is discrete, as it only stands for the indexing of sequences and does not have any temporal meaning, in dynamic time warping it may be continuous. That means that sequences ${\bf a}$
and ${\bf b}$ are in fact continuous trajectories, $\{a(t)\}_{t\in \T}$ and
$\{b(s)\}_{s\in \T}$, taking generally real values. In this work we will
combine both approaches. Indeed, in the functional data context, the observed curves
are real-valued and have a continuous nature. However, we will
focus on event type data.\fi

In our setting of event data, we 
observe pairs of curves ${\bf a}= (Y_i(t_{i1}), \ldots, Y_i(t_{in_i}))$ and ${\bf
b}=(Y_j(t_{j1},\ldots,Y_j(t_{jn_j})$  at event times ${\bf t}= (t_{i1}, \ldots, t_{in_i})$ and ${\bf
s}=(t_{j1},\ldots,t_{jn_j})$, respectively. Then, to  align curves $Y_i$ and $Y_j$ 
one needs to determine, first, which elements of the sequence ${\bf a}$  should  be mapped to which
elements of the sequence ${\bf b}$, 
 and second,  the time points at which these aligned values should
to be placed. For the first determination, a discrete DTW approach, which is described
next, is used. The second determination is addressed in Section \ref{algo}.

From now on and in the following, we use the notation ${\bf a}=(a_1, \ldots, a_{\ell})$ and ${\bf b}=(b_1,\ldots,b_m)$ to refer to the observed values, without any temporal reference, of any pair of curves. We now describe the DTW algorithm that allows us to find an optimal mapping between these two sequences.

Let $d: {\cal A} \times {\cal A} \rightarrow \R$ be a suitable distance
function defined on the feature space. Then, for any alignment $\e$, its
alignment distance is given by
\begin{equation}\label{dis}
D_{\e}({{\bf a}},{{\bf b}})=\sum_{k=1}^{|\e|} w(k)\cdot d(a_{L_k}, b_{M_k}),
\end{equation}
where $w(\cdot)$ represents time weights that account for the length
of the alignment. Different definitions of these weights are possible, even $w(\cdot)=1$ (see \cite{wang:97} for the study of weight functions in the continuous time DTW problem).
In this work we consider the definition given by \cite{KruskalLib}, $w(k)=(t_{L_k} - t_{L_{k-1}} + s_{M_k}
-s_{M_{k-1}})/2$, which is quite standard in the DTW literature.

Then, the optimal alignment is
\begin{equation}\label{opt_alig}
\e^{ab}= \argmin_{\e \in \EE_{{\ell},m}} D_{\e}({{\bf a}},{{\bf b}}),
\end{equation}
where $\EE_{{\ell},m}$ is the set of all possible alignments of sequences with
lengths ${\ell}$ and $m$. The dynamic time-warping algorithm used to retrieve
$\e_{ab}$ is as follows:

\begin{enumerate}
\item Initialize:\\
$D_{(1,1)}= 0, \quad I_{(1,1)}=(1,1),$\\
$D_{(i,1)}=D_{i-1,1} + d(a_i,b_1)\cdot (t_i - t_{i-1})/2 , \quad I_{(i,1)}=(1,0),\quad i=2,\ldots,{\ell},$\\
$D_{(1,j)}=D_{1,j-1} + d(a_1,b_j)\cdot (s_{j} -s_{j-1})/2, \quad
I_{(1,j)}=(0,1),\quad j=2,\ldots,m.$
\item Iterate for $i=2,\ldots,{\ell}$, $j=2,\ldots,m$:\end{enumerate}
$$D_{(i,j)}\!\!= \!\! \min \!\! \left\{ \begin{array}{cll}
\!\! D_{(i-1,j-1)}\!\! \!\! &+& \!\!\!d(a_i,b_j) \!\cdot\! \dfrac{(t_i - t_{i-1} + s_{j} -s_{j-1})}{2} \vspace{0.1cm}\\
\!\! D_{(i-1,j)}  \!\! \!\! &+& \!\!\!d(a_i,b_j) \!\cdot\! \dfrac{(t_i - t_{i-1})}{2} \!\cdot \!\1{I_{(i-1,j)}\!\!\neq\!\!(0,1)} \!+\! M \1{I_{(i-1,j)}\!\!=\!\!(0,1)}\vspace{0.1cm}\\
\!\! D_{(i,j-1)}  \!\!\!\!  &+& \!\!\!d(a_i,b_j) \!\cdot \!\dfrac{(s_{j} -s_{j-1})}{2}\!\cdot\! \1{I_{(j-1,i)}\!\!\neq\!\!(1,0)}\!+\! M \1{I_{(j-1,i)}\!\!=\!\!(1,0)}
 \end{array}\right.  $$
$$
I_{(i,j)}\!\!=\!\!  \left\{ \begin{array}{cll}
(1,1) &\mbox{ if }& D_{(i,j)}=D_{(i-1,j-1)} + d(a_i,b_j) \cdot \dfrac{(t_i - t_{i-1} + s_{j} -s_{j-1})}{2}\vspace{0.1cm}\\
(1,0) &\mbox{ if }& D_{(i,j)}=D_{(i-1,j)}   + d(a_i,b_j) \cdot \dfrac{(t_i - t_{i-1})}{2} \vspace{0.1cm}\\
(0,1) &\mbox{ if }& D_{(i,j)}=D_{(i,j-1)}   + d(a_i,b_j) \cdot \dfrac{(s_{j} -s_{j-1})}{2}
 \end{array}\right.  $$
\begin{enumerate}[resume]
\item $d_{\e^{ab}}({{\bf a}},{{\bf b}})=D_{({\ell},m)}.$
\item Trace back:\\
$\e^{ab}_{|\e^{ab}|}=I_{({\ell},m)}$, $i={\ell}-I_{({\ell},m)}(1), j=m-I_{({\ell},m)}(2)$,
$k=|\e^{ab}|$ and while $i\geq 1$ and $j\geq 1$ do:
$$\e^{ab}_{k}=I_{(i,j)}, \quad k=k-1,$$
\begin{equation}\label{DTW}
i=i-I_{(i,j)}(1), \quad j=j-I_{(i,j)}(2).
\end{equation}
\end{enumerate}
Note that in step 1, the initialization $D_{(1,1)}=
0$, $I_{(1,1)}=(1,1)$, forces the two first values of the two sequences to be
aligned together. In step 2, we prevent having a vertical movement followed
by an horizontal one, or vice versa, by choosing a sufficiently large
constant $M$. This means that if for instance $a_i,\ldots,a_{i+k}$, $k>0$,
are all mapped to $b_j$, then $b_{j+1}$ can not be mapped again to $a_{i+k}$. As for the computational cost, it is ${\cal O} ({\ell}\cdot m)$, which is low, since ${\ell}$ and $m$, the number of events per curve, are usually quite small.

Once we obtain the optimal alignment, $\e^{ab}$, of sequences {{\bf a}} and {{\bf b}}, what we get is a correspondence between values $a_1,\ldots,a_{\ell}$ and $b_1,\ldots,b_m$, without any timing reference. That is, at this stage we do not tackle the problem of designating the time points at which two (or more) aligned values should be placed. This is a major difference between our approach and  previous work using DTW related algorithms in the context of time warping, as for example \cite{wang:97, wang:99}. Indeed, these works deal with continuous features and time synchronization problems in which finding similarities between points in different curves and assigning temporal references cannot be tackled separately. In \cite{Bigot:06}, a dynamic programming algorithm similar to (\ref{DTW}) is used to find a correspondence between two previously identified sets of landmarks from two curves. Then, a penalized least squares minimization problem is solved to estimate the warping functions that register the two curves through the alignment of their landmarks.

Our approach is quite different since we focus on global alignment of the whole set of curves for which pairwise alignment maps constitute  just a preliminary stage. That is, we use DTW to determine the pairwise correspondence of events, and this information about pairwise correspondences is then combined (see Section \ref{algo}) to
obtain global alignment, i.e., to register the entire sample. This is a key difference, as the main disadvantage of DTW based registration methods is that the generalization from two to a larger number of curves is not straightforward. Indeed, extensions to the alignment of a larger set of curves usually lie in defining a common set of landmarks (from individual landmark vectors or from a template curve) to which then the entire sample of curves is aligned. However, this ad hoc approach suffers from a loss of information and requires computationally intensive algorithms.

In the last decade, DTW has been used for the alignment of
expression profiles from time-course microarray experiments (see \cite{Aach}
and \cite{Clote}). In this context, a $n$-dimensional time series,
containing the expression levels of a group of $n$ genes recorded at the same
time points (and which are assumed to be synchronized), is aligned to another $n$-dimensional time series corresponding to a different microarray experiment of the same $n$ genes. That is, DTW is used
to perform pairwise alignment of two $n$-dimensional sequences. This is different from using DTW to align $n$ different trajectories. Indeed, although multiple dynamic time warping is theoretically possible and algorithmically
equivalent to the pairwise case, it is computationally unfeasible even for a
relatively small number of sequences, since the computing time is proportional to
$\ell^n$, where $n$ represents here the total number of sequences and $\ell$ their
length, assuming they all have similar lengths.

\subsection{Pairwise dynamic time warping}\label{algo} The proposed strategy is to
combine DTW, to obtain pairwise alignment maps in a first step,
with pairwise registration. The goal is to extract information about global alignment from the pairwise correspondences.

Then, given a sample of $n$ curves, we aim at estimating $\hat{g}_{ij}(t)$
and $\hat{g}_{ji}(t)$ for any $i,j=1,\ldots,n$. The observations in
model (\ref{tw}) are $(t_{ik}, y_{ik})$, $k=1,\ldots,n_i$, $i=1,\ldots,n$, where $t_{ik}$ is the time at which the $k$-th event is observed for individual $i$ and $y_{ik}$ is the observed value of $Y_i$ at time $t_{ik}$. 

However, in what follows we will focus on any given pair of curves $y_i$ and $y_j$, and for the sake of simplicity  will denote the corresponding observations as $(t_{k}, y_{ik})$, $k=1,\ldots,n_i$ and $(s_k,
y_{jk})$, $k=1,\ldots,n_j$. Applying algorithm (\ref{DTW}) we obtain an optimal
mapping, $\e^{ij}$, of the values $y_{i}$ and $y_{j}$. But one thing is to
decide which values on both sequences can be aligned together, and another
thing is to decide the time points corresponding to the aligned records.
However, recall that in our case the alignment of pairs of curves is
just a preliminary step to arrive at a global alignment of the whole sample of curves
and is not an objective by itself. From estimated pairwise alignments 
$\hat{g}_{ji}(t)$ and $\hat{g}_{ij}(t)$, we also obtain
$\hat{g}_{ji}(t_k)=s_h$ and $\hat{g}_{ij}(s_h)=t_k$. However, if more than one value in one curve is mapped to the same single value in the other curve some additional considerations
are needed.

\begin{figure}[h!]
\centering
\includegraphics[trim=75mm 3mm 62mm 0mm, clip,width=15cm]{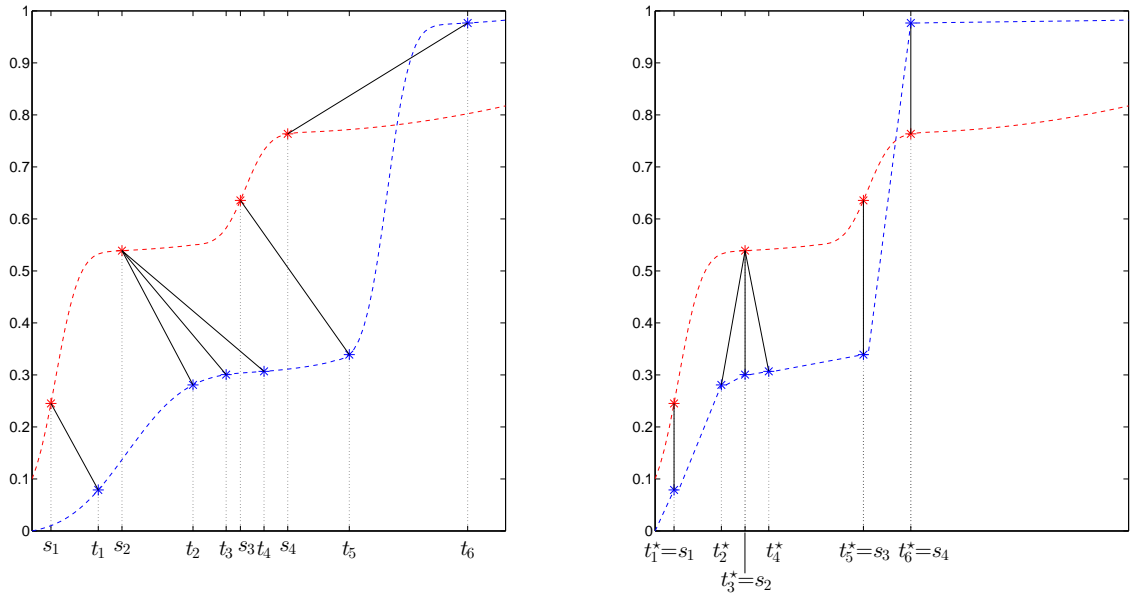}
\caption{Left: Blue stars are points $(t_k,y_{ik})$ and red stars are points $(s_k,y_{jk})$.
Solid black lines represent the alignment obtained by DTW. Imaginary
continuous trajectories are displayed in dashed lines, although these
are not observed in practice. Right: Alignment of $y_i$ towards $y_j$.
 Blue stars are now points $(t^{\star}_k=\hat{g}_{ji}(t_k),y_{ik})$. Points
 $(t_2,y_{i2})$, $(t_3,y_{i3})$, $(t_4,y_{i4})$ although all mapped to $(s_2,y_{j2})$, are spread in a small interval around $s_2$. }\label{Finterp}
\end{figure}

We now describe the detailed procedure to obtain $\hat{g}_{ij}(t)$ and
$\hat{g}_{ji}(t)$ from observed sequences  $(t_k, y_{ik})$, $k=1,\ldots,n_i$,
and $(s_k, y_{jk})$, $k=1,\ldots,n_j$:
\begin{itemize}
\item[-] Find $\e^{ij}$, the optimal alignment between the two sequences
    applying algorithm (\ref{DTW}) with $d$ being the Euclidean distance
    in $\R$. Denote $p= |\e^{ij}|$.
\item[-] $(L_{1:p}, M_{1:p})=\left\{ \sum_{k=1}^r \e^{ij}_k
    \right\}_{r=1,\ldots,p}$.
\item[-] $\alpha_1=M_1$ and $\alpha_{2:n_i}=\{M_s, s=2,\ldots,\ell, \mbox{ s.t. } L_s\neq L_{s-1} \}$\\
$\beta_1=L_1$ and $\beta_{2:n_j}=\{L_s, s=2,\ldots,\ell, \mbox{ s.t. }
M_s\neq M_{s-1} \}$.
\item[-] For $k=1\ldots,n_i$:\end{itemize}
\begin{equation}\label{tstar}\begin{array}{llclll}
\hspace{0.4cm}&t^{\star}_k\!\!\!\!&=&\!\!\!s_{\alpha_k} &&\mbox{if } \alpha_{k-1}\! \neq \!\alpha_k \!\neq\!
\alpha_{k+1} \mbox{ (only one of them if $k=1$ or $n_i$)}\\
&t^{\star}_{k+h}\!\!\!\!&=&\!\!\!s_{\alpha_k}\!-\! l\!+\!\dfrac{2l}{H-1}h, &&  h=0,\ldots,H-1,
\,\, \mbox{ with } l=\dfrac{\delta(t_{k+H-1}-t_k)}{2}\\ &&&&& \mbox{ if }
\alpha_{k-1}\neq \alpha_{k}= \alpha_{k+1}=\cdots=\alpha_{k+H-1} \neq \alpha_{k+H}
\end{array}
\end{equation}
\begin{itemize}
\item[-] For $k=1\ldots,n_j$:\end{itemize}
\begin{equation}\label{sstar}\begin{array}{llclll}
\hspace{0.4cm}&s^{\star}_k\!\!\!\!&=&\!\!\!t_{\beta_k} &&\mbox{if } \beta_{k-1} \!\neq \!\beta_k \!\neq \!
\beta_{k+1} \mbox{ (only one of them if $k=1$ or $n_j$)}\\
&s^{\star}_{k+h}\!\!\!\!&=&\!\!\!t_{\beta_k}\!-\! l\!+\!\dfrac{2l}{H-1}h, &&h=0,\ldots,H-1, \,\,
\mbox{ with } l=\dfrac{\delta(s_{k+H-1}-s_k)}{2}\\ &&&&&
\mbox{ if } \beta_{k-1}\neq \beta_{k}= \beta_{k+1}=\cdots=\beta_{k+H-1} \neq \beta_{k+H}
\end{array}
\end{equation}
\begin{itemize}
\item[-] Define
\begin{equation}\label{interp}
\hat{g}_{ji}(t_k)=t^{\star}_k,\,\, k=1,\ldots, n_i \quad \hat{g}_{ij}(s_k)=s^{\star}_k,\,\, k=1,\ldots, n_j.
\end{equation}
\end{itemize}
Note that expressions (\ref{tstar}) and (\ref{sstar}) correspond to an
interpolation step in the cases in which more than one value of $y_i$ has
been assigned to the same single value of $y_j$ and vice versa.
In those cases, we spread those values in a certain interval rather than
bringing them together to the same time point. That is, if values
$y_{ik},\ldots, y_{i,{k+H-1}}$ are all mapped to $y_{j,\alpha_k}$, instead of
defining $\hat{g}_{ji}(t_{k+h})=s_{\alpha_k}$, $h=0,\ldots,H-1$, we equally
distribute these points around $w_{\alpha_k}$ in such a way that the slope of
$\hat{g}_{ji}$ between $t_{k}$ and $t_{k+H-1}$ is equal to $\delta$ (where
$\delta$ is a parameter that needs to be chosen in advance). In this way we
guarantee that the estimates $\hat{g}_{ji}(t)$ and $\hat{g}_{ij}(t)$ are
strictly increasing. This is illustrated in Figure \ref{Finterp}.

Once we have discrete time estimates of $g_{ji}(t)$ and $g_{ij}(t)$, for any
pair $i,j=1\ldots,n$, $i\neq j$, following \cite{TangMuller_Bio} we define:
\begin{equation}
\hat{h}_i^{-1}(t_k)=\dfrac{1}{n-1} \sum_{j\neq i}  \hat{g}_{ji}(t_k), \quad k=1,\dots, n_i, \quad i=1,\ldots,n.
\end{equation}
We note here  that each estimated warping function, $\hat{h}_i^{-1}$, is only
defined over the grid of observed event times of its corresponding curve,
$y_i$. The same applies for the estimated registered curves $\hat{x}_i=y_i
\circ \hat{h}_i$. Then, any further step involving computations with either
the estimated warping functions or registered curves, such us calculating the
sample mean of the registered curves for instance, requires an additional
interpolation step aiming to anchor curves at a common grid of time points.

As for the computational complexity of the pairwise dynamic time warping
algorithm to produce an alignment of a set of $n$ curves, this is
proportional to $\binom{n}{2} {\ell}^2 \approx (n\cdot \ell)^2$, ($\ell$ standing for
some average number of events). This represents a very important reduction
compared to the dynamic time warping approach for multiple alignment, which
has a computational cost of ${\cal O}({\ell}^n)$, as we have already mentioned.
\section{Applications}\label{app}

\subsection{Historical fertility data}\label{SFert}
We study fertility from a well-documented 17/18th century cohort of
1877 native born French-Canadian women who lived past age 50. This
data contain the number of children and the age at the different
births for each woman in the sample. We consider only those women with
more than one child; the resulting data  consist of
1810 individuals. For more details on this data set see \cite{Fertil} and the references therein.

\begin{figure}
\centering
\includegraphics[width=15cm]{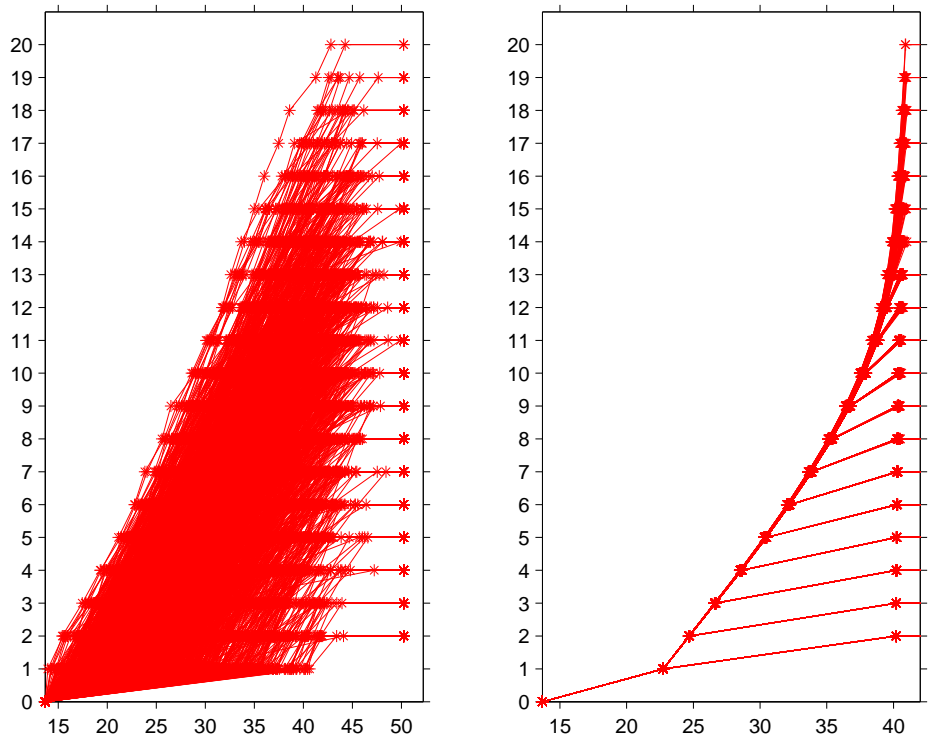}
\caption{Left: Absolute number of children along time for 1810 women. Stars represent the observed data,
solid lines are obtained by connecting births within each record. Right:  Absolute number of children versus registered births times for 1810 women.}\label{Fert1}
\end{figure}

\begin{figure}
\centering
\includegraphics[width=15cm,height=10cm]{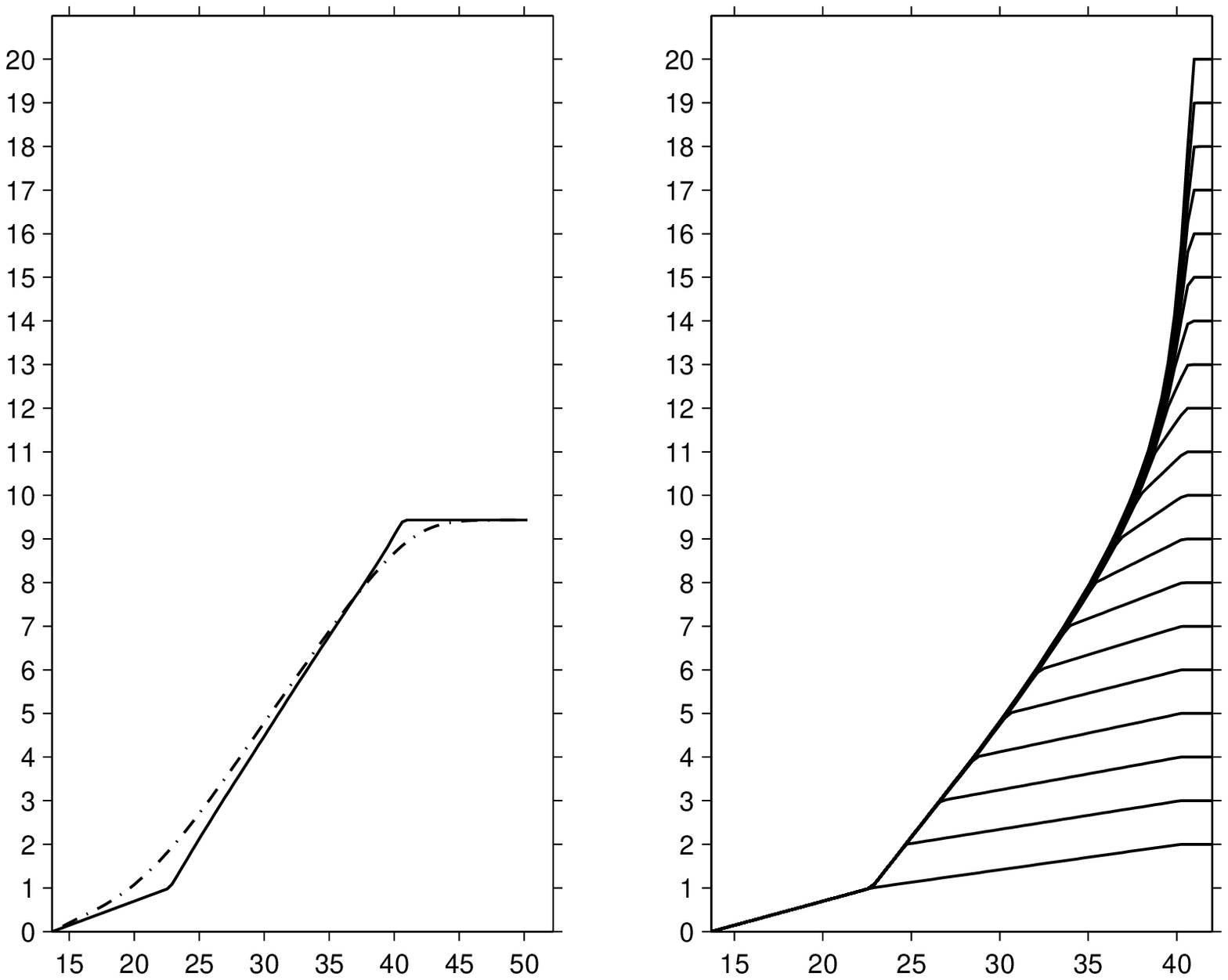}
\caption{Left: Mean intensity function before (dashed-dotted line) and after registration (solid line). Right: Group mean intensity functions. The groups are defined by the women with the same number of children.}\label{Fert_means}
\end{figure}

To study the relation between the total number of children and the
dynamics of the giving birth process, we transform the data to
obtain a curve $(t_{ik}, y_{ik})$, $i=1, \ldots, n_i$ for each woman, where $t_{ik}$ is the age
of the $i$-th woman at her $k$-th birth, $y_{ik}=k$ and $n_i$ is
 her total number of children. The time domain is $\T=[13.7, 50.2)$, units are years. To
   anchor all sample curves on this interval, we add points $(13.7, 0)$ and $(50.2,n_i)$ at the beginning and at the end of each curve. Since this introduces an  artificial constant fragment at the end of each curve an  important reference  is the age at the last birth for each woman and not the end of the interval, we force points $(t_{in_i},y_{in_i})$, $i=1,\ldots,n$, to be mapped together. The whole sample before and after registration by the procedure described in Section \ref{algo} is
presented in Figure \ref{Fert1}. Note that the registered sample includes contracted time domains for some women so that the periods between births may be shorter than 9 months. This is a consequence of the forced alignment
at the first and last births across  all women and is also reflected in some of the mean group intensity functions (averages of the registered curves for women with the same number of offspring) displayed in Figure \ref{Fert_means}. Nevertheless, the mean intensity function for the whole sample presents a good approximation of the standard fertility process and captures a linear trend along with the mean ages at the first and last birth, which are 22.4 and 40.2 years respectively. In the case of the sample mean of the original trajectories (Figure \ref{Fert_means}), these ages are under-, respectively,  
over-estimated.
After registration, the estimated warping
functions can be used for classification purposes. We consider a distance-based approach by defining individual distances from the warping functions as follows, 
\begin{equation}\label{dist_ind}
d(i,j)=\int_{\T} \left(\hat{h}^{-1}_i (t) - \hat{h}^{-1}_j (t) \right)^2 dt, \quad i,j=1\ldots,n.
\end{equation}
Indeed, since the warping functions are strictly increasing and defined from $\T$ to $\T$, the area between two of these functions is a good indicator of the difference in the degree of time distortion that they introduce in the respective individual curves. We expect that this distance will be useful at discriminating between different fertility dynamics. This distance matrix needs to be computed numerically from the discrete-valued warping
function estimates. 

Unlike common approaches in functional data analysis, we propose to perform clustering directly from the distance matrix without any previous step to reduce the dimensionality of the data. The clustering method is a modified version of the k-means algorithm in which the centroid of a group $G$, is calculated as its Fr\'{e}chet mean, $c_G=\arg \min_{x\in G} \sum_{y\in G} d^2(x,y)$.
To choose the number of clusters we use the \emph{silhouette} coefficient \citep{Kauf_Rous_90} that for each individual provides a measure of how well classified it is. The \emph{silhouette} of individual $i$ is $(b_i-a_i)/\max\{a_i,b_i\}$, where $a_i$ is the mean distance of $i$ to the rest of individuals in the same cluster and $b_i$ is the mean distance of $i$ to the individuals of the closest cluster (besides the one it is assigned to). Finally, the \emph{silhouette} coefficient of a clustering is the mean \emph{silhouette} across individuals. Then, the choice of the number of clusters is done by maximizing the \emph{silhouette} coefficient. For this data set, the maximum was found for $k=2$ clusters, with a value of $0.60$. According to \cite{Kauf_Rous_90}, a value of this coefficient between $0.51$ and $0.70$ indicates that a reasonable structure has been found in the data.
 
\begin{figure}
\centering
\includegraphics[width=15cm]{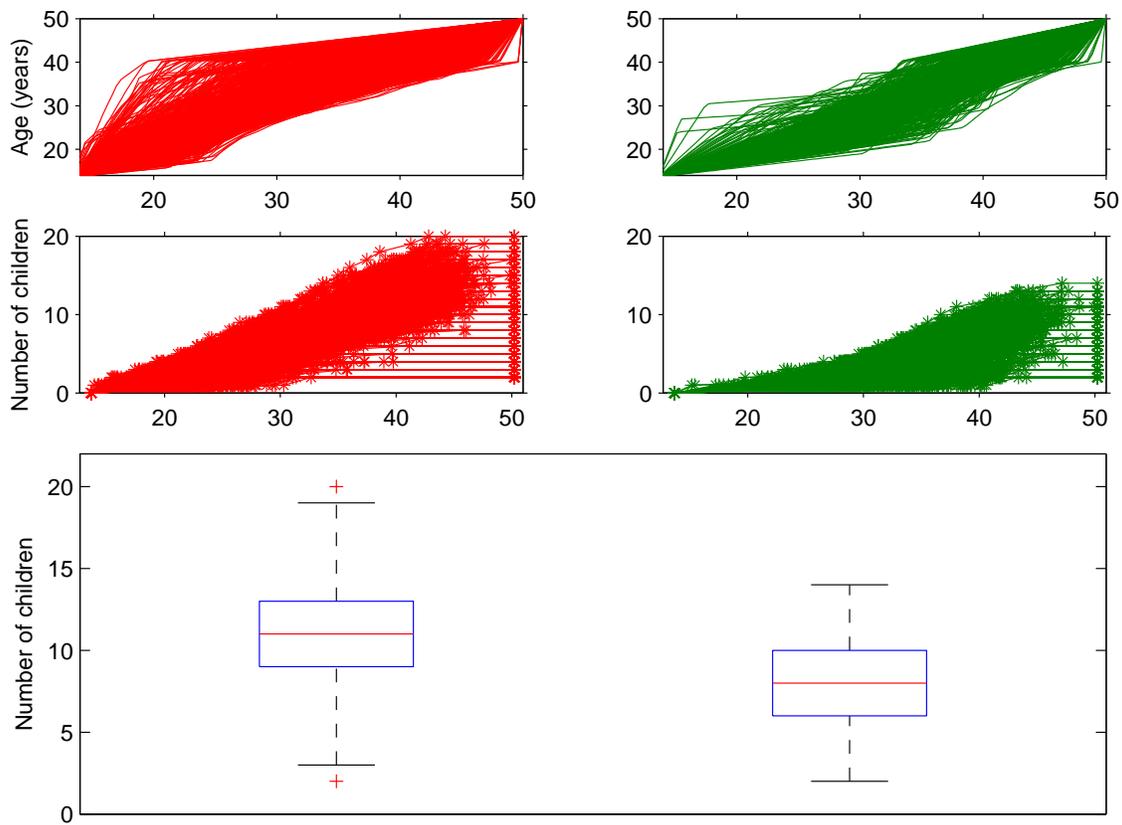}
\caption{Warping function estimates ($\hat{h}_i^{-1}$), fertility curves and box plots of the number of children for the two clusters.}\label{Fert3}
\end{figure}

The results for $k=2$ clusters are displayed in Figure \ref{Fert3}. The two clusters can be defined as women with late birth trajectories, and women with regular birth trajectories. A regular birth trajectory is almost linear, the first birth occurring in average at 20.8 years and the last birth at 39.7 years. In contrast, a late birth trajectory is piecewise linear with two differentiated pieces, the first one with lower slope than the second one. In fact the slope of the second fragment is similar to that of a regular trajectory. The average ages at the first and last birth for women in this category are 29.4 and 41.8 years. Then, we can say that the main difference between the women in the two clusters is the age at which the first child is born, since after the first birth, the fertility process looks similar in both cases.
If we now look at the distribution of the number of children in these two clusters we observe an expected consequence, namely, that the total number of children is generally lower for those women that had their first child at an older age.

\subsection{Online auction data}\label{SAuction}
Modelling of price paths in on-line auction data
has received a lot of attention in recent years
\citep{AD1, AD2, AD3, AD4}. One of the reasons is
the availability of huge amounts of data made
public by the on-line auction and shopping website
eBay.com, which has become a global market place in
which millions of people worldwide buy and sell
products. The price evolution during an auction can
be thought as a continuous process which is
observed discretely and sparsely only at the
instants in which bids are placed. In fact, bids
tend to concentrate at the beginning and at the end
of the auction, corresponding to two typically
observed phenomena, ``early bidding'' and ``bid
sniping'' (a situation in which ``snipers'' place
their bids at the very last moment). Here, we
analyze a set of 158 eBay auctions for Palm M515
Personal Digital Assistants (PDA), of a fixed
duration of seven days, that took place between
March and May, 2003. This data set is publicly
available at
\text{http://www.rhsmith.umd.edu/digits/statistics/data.aspx}
and has been previously studied by \cite{AD1} and
\cite{AD4}, among others. As in \cite{AD4} we
restrict our analysis to 156 auctions after
removing two irregular recordings.

Our interest here is not focused on the price process, but rather on the intensity functions that quantify the bid arrivals along the time axis, where each
bid arrival generates an event time. This approach may be useful to understand different bidding behaviors and for characterizing the bid arrivals process. A characteristic of these data is that bidding trajectories for different
auctions are quite dissimilar. While most of them reflect very low bidding activity at the
beginning of the auction and intense bidding when
the auction is near its end, in other auctions one observes different patterns. 
It seems reasonable to view these
trajectories as accelerated and decelerated versions
of an underlying bidding intensity function and therefore the proposed time warping model seems
suitable for this analysis.

\begin{figure}
\centering
\includegraphics[width=15cm,height=10cm]{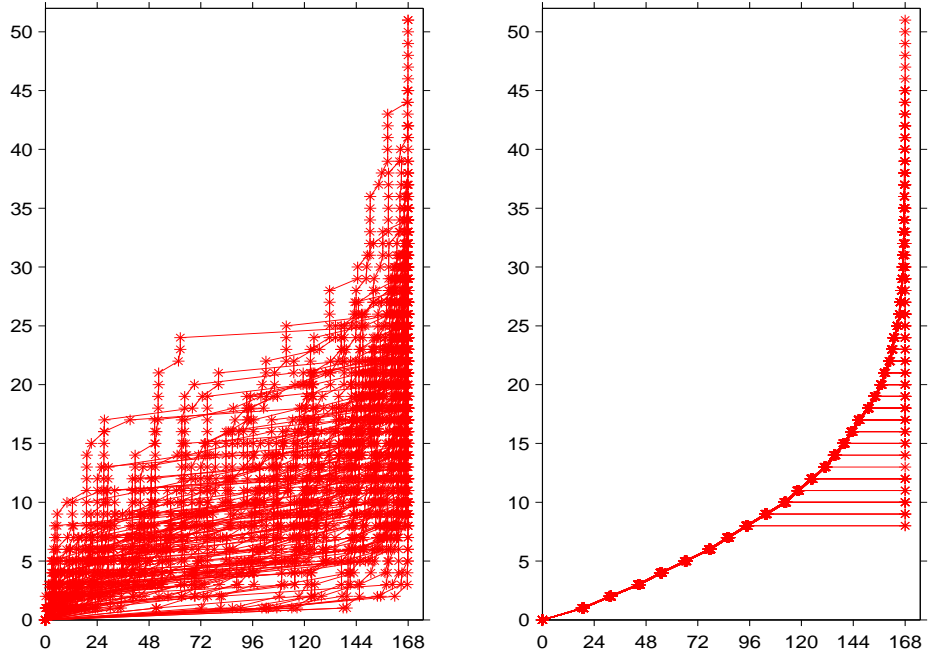}
\caption{Left: Observed bidding trajectories $Y_i$ for 156 auctions.
Stars represent the observed data, solid lines are
connecting the stars. Right:
Registered bidding trajectories.}\label{Auction1}
\end{figure}

\begin{figure}
\centering
\includegraphics[width=15cm,height=10cm]{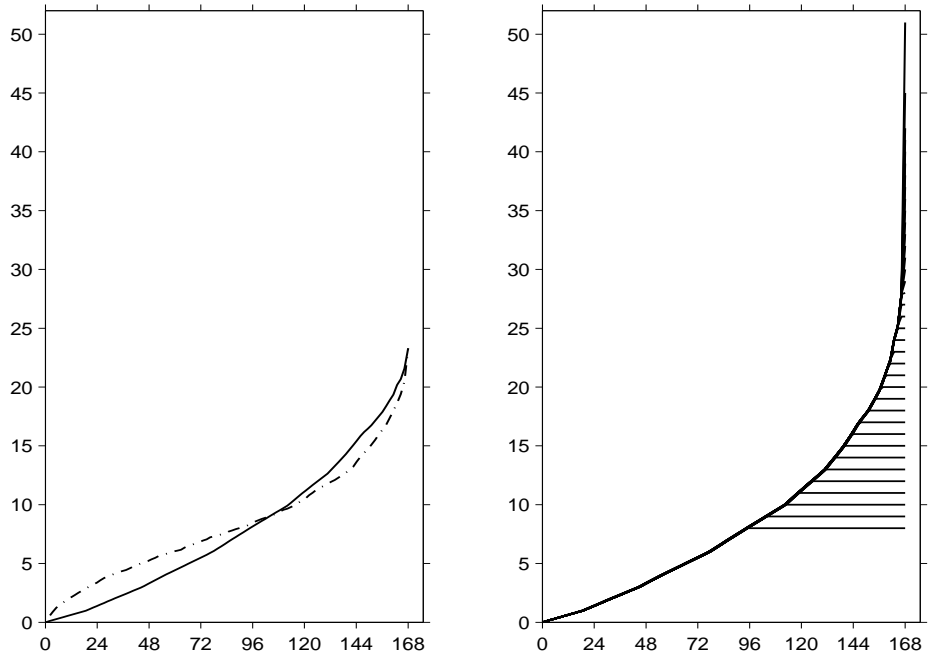}
\caption{Left: Mean intensity function before (dashed-dotted line) and after registration (solid line). Right: Group mean intensity functions. The grouping is defined by the number of bids.}\label{Auction_means}
\end{figure}

The time domain for observations is $\T=[0, 168)$, where the time units are
hours. As in the previous example, we have added points $(0,0)$ and $(168,n_i)$ at the beginning and at the end of each trajectory. The minimum and maximum observed numbers of bids for auctions in the sample are 8 and 51 respectively. Applying  the algorithm described in Section \ref{algo}, with  results  shown in
Figure \ref{Auction1}, indicates that the registered (non-standardized) intensity functions
show a clear exponential-type  pattern. We obtain an ``almost common'' intensity function irrespective of the number of bids. Indeed, the function is the same for all auctions except for the time period after the last bid, which is represented by a constant straight line at a height that depends on the total number of bids.  This time period has been expanded (resp. contracted) during the registration process in those auctions with a low (resp. high) number of bids. Due to the low variability in the registered sample, the group mean intensity functions (Figure \ref{Auction_means})  look similar to the registered trajectories. Also in Figure \ref{Auction_means}, the overall mean intensity function after registration is compared to the sample mean of the observed curves. We can appreciate how  ``early bidding'' and ``bid sniping''  emerge as
phenomena that are due to time warping of the overall mean intensity function. 

\begin{figure}
\centering
\includegraphics[width=15cm]{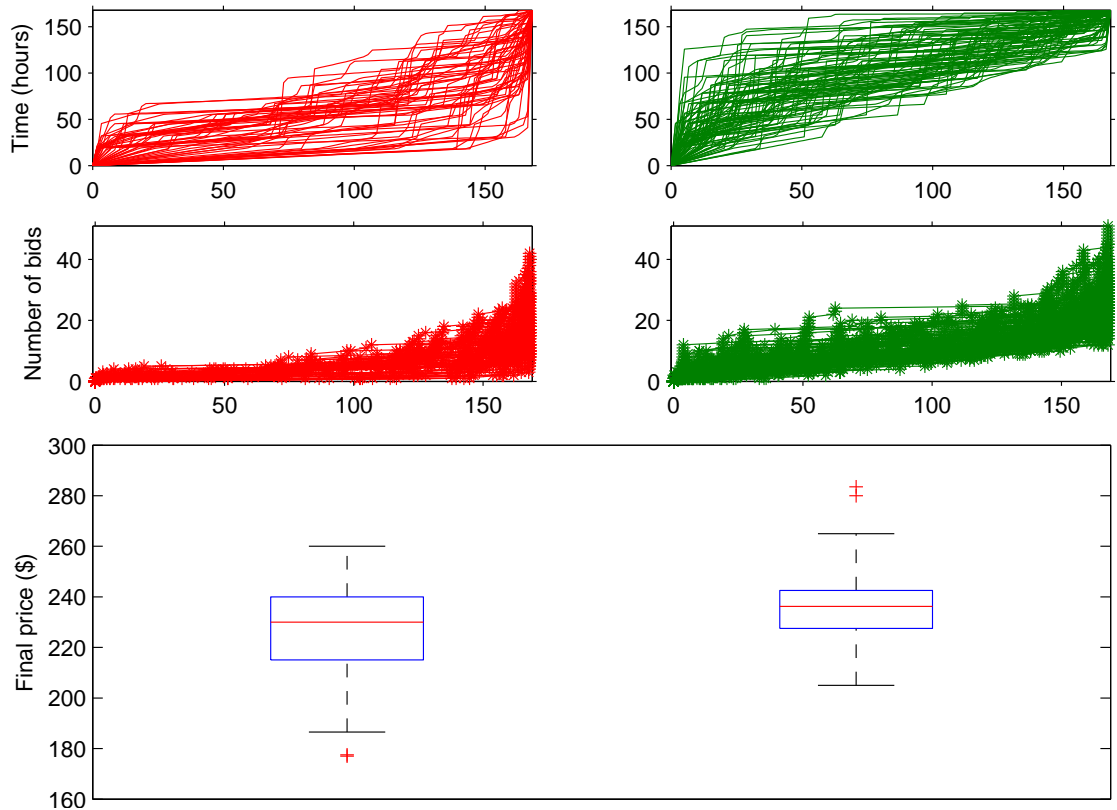}
\caption{Warping function estimates ($\hat{h}_i^{-1}$), bidding trajectories and box plots of the final auction price for the two clusters ``L'' (left) and ``R-E'' (right).}\label{Auction3}
\end{figure}

Aiming to differentiate between  different bidding behaviors, we apply the warping-based
clustering procedure described in Section
\ref{SFert}. Here, again the number of clusters was found to be $k=2$, with a \emph{silhouette} coefficient of $0.65$.
In Figure \ref{Auction3} we present the 
warping function estimates, bidding trajectories and final price distributions
of the two clusters. In fact, we want to investigate any possible relation between the bidding dynamics and the closing price of an auction. The clusters are defined by auctions with late (``L'') and regular-early (``R-E'') bidding activity. This is obvious if we look at the warping function estimates corresponding to each group. However, it is more difficult to tell the differences between the observed bidding trajectories of the two clusters. Some of them, although classified in different groups may be close together in terms of an $L^2$ distance on the space of observed trajectories. Then, the warping-based clustering provides a useful tool for understanding bidding activity. The first cluster, ``L'', is composed by auctions with a very slow start, and contains most of the auctions with lowest final prices. Indeed the median final price of cluster ``L'' is slightly lower than that of ``R-E''. However it also contains auctions with high closing prices (ranging between 240 and 260 \$), although these systematically correspond to auctions with unusually high opening bids (ranging between 50 and 179.99 \$). The second cluster, ``R-E'', contains both auctions with bidding trajectories that are similar to the mean intensity function obtained after registration and auctions presenting very intense early bidding activity. The bidding trajectories of the two groups seem to present a similar final shape (a very important increase in the bidding activity during the last day of the auction) up to a vertical shift. That is, bidders behavior during the last day of an auction seems to be the same independently of the previous bidding activity until that moment. According to this, one could think that the first hours of an auction determine its outcome. However, this may not be completely true. In fact, in cluster ``R-E'' one can distinguish two very different starting behaviors, which do not correspond to different price distributions. Indeed, if one chooses to divide the data into 3 groups, cluster ``R-E'' is split into an ``R'' and an ``E'' cluster for which the closing price distributions are very similar. As a conclusion, we could say that it is a very low activity during the first 3/4 of an auction that yields to lower closing prices. In contrast, auctions with an early intense bidding activity do not necessary lead to higher final prices than those with a regular starting.

\section{Discussion}
We have proposed an efficient registration algorithm for functional data that can be characterized by 
a random number of event times, which could correspond to original observations, or alternatively
 could be derived features from functional data such as the location of peaks, which are not consistently observed across the random trajectories that constitute the functional data sample. One of the advantages of the proposed method is that it can cope with large sets of curves.  Unlike previous DTW-based strategies for continuous time curve registration we focus on discrete time pairwise ``local'' alignments that subsequently are combined to obtain a whole sample ``global" alignment.

The method relies on the choice of a metric on the observations space, which is used to find matching values over different curves. Throughout this article we consider the Euclidean distance for this purpose, although any other metric could be used. Indeed, this is a crucial point: different metrics will provide different alignments. Knowing which values should be aligned together is equivalent to having a notion of what similarity means in each particular application. For example, in the applications of Section \ref{app}, we focus on aligning some particular events, such as the birth of the first child, for the fertility data. For this purpose, the Euclidean distance proves to be quite suitable. 

As for the clustering of a sample of curves, one may wonder why the distance-based method described in Section \ref{SFert} is not directly applied to the observed curves instead of the warping estimates.
 Of course this could be done, but the results will be different, due to differences in the distance matrix used for clustering. Here again, different similarity notions can be suitable for different purposes. In this context, our proposal of clustering based on warping estimates can be seen as a way of defining a distance between individuals which focuses on similar temporal behavior.

\section*{Acknowledgments}
This research was supported by NSF grants DMS-1104426 and DMS-1228369, and by Spanish grants ``Jos\'{e} Castillejo'' JC2010-0057, MTM2010-17323 and ECO2011-25706 (Spanish Ministry of Science and Innovation).

\bibliographystyle{plainnat}
\bibliography{bib_PDTW}

\end{document}

\end{document}